\documentclass[reprint,aps,prb]{revtex4-1} 

\usepackage{amsmath}
\usepackage{graphicx}

\begin{document}

\title{Do the standard expressions for the electromagnetic field-momentum need any modifications?}
\author{Ashok K. Singal}
\email{asingal@prl.res.in}
\affiliation{Astronomy and Astrophysics Division, Physical Research Laboratory,
Navrangpura, Ahmedabad - 380 009, India}

%\date{\today}

\begin{abstract}
We investigate here the question raised in literature about the correct expression for the electromagnetic field-momentum, 
especially when static fields are involved. For this we examine a couple of simple but intriguing cases. First we consider a system 
configuration in which electromagnetic field momentum is present even though the system is static. 
We trace the electromagnetic momentum to be present in the form of a continuous transport of electromagnetic energy from one part of 
the system to another, without causing any net change in the energy of the system. 
In a second case we show that the electromagnetic momentum is nil irrespective of whether the charged system is stationary or in motion, 
even though the electromagnetic energy is present throughout. 
We demonstrate that the conventional formulation of electromagnetic field-momentum describes the systems consistently without any real 
contradictions. Here we also make exposition of a curiosity where electromagnetic energy decreases when the charged system gains velocity.  
Then we discuss the more general question that has been raised -- Are the conventional formulas for energy-momentum of electromagnetic fields 
valid for all cases? Or more specifically, in the case of so-called ``bound fields,'' do we need to change over to some modified definitions  
which have  made their appearance even in standard text books? We show that in all cases it is only the conventional formulas which lead to 
results consistent with the rest of physics, including the special theory of relativity, and that any proposed modifications are thus 
superfluous.
\end{abstract}
\maketitle
%---------------------------------------------
\section{Introduction}
A question about the correct expression for the electromagnetic field momentum, especially for static cases, 
has been raised in the literature. \cite{Romer95, Romer66, Griffiths12} The standard expression for the electromagnetic field momentum is 
the volume integral,\cite{1,25}
\begin{equation}
\label{eq:p31.1}
{\bf  P}=\epsilon_{\rm o}\int_{\rm V}{\bf E} \times {\bf B}\:{\rm d}v. 
\end{equation}
The expression is believed to be true even for static systems where the sources of the electric and magnetic fields may not be 
having any apparent motion, with the fields thereby being static with no temporal change in the field values.
For instance, in a static system comprising a pair of 
crossed electric and magnetic dipoles (${\bf p}$ and ${\bf m}$ respectively), 
there is a linear electromagnetic momentum, ${\bf  P} \propto {\bf m} \times {\bf p}$, associated with the system. 
The puzzling question then is -- what really is moving in this otherwise static system? 
The momentum is certainly not due to the drift velocities of charges that constitutes the steady current giving 
rise to ${\bf m}$; after all that motion is present even if the electric dipole 
${\bf p}$ were absent and in that case there would not be any electromagnetic momentum ${\bf P}$ in the system. Even otherwise, any 
such momentum vector due to drift velocities of conducting charges, when integrated over a closed current loop, 
will have a nil value. Nor could the electric dipole ${\bf p}$ alone be the cause of any momentum ${\bf  P}$. Electromagnetic field momentum 
makes an appearance only when both ${\bf p}$ and ${\bf m}$ are present and that too when they have at least some components mutually 
perpendicular. For further details and references on the question of electromagnetic momentum we refer the reader to the 
review article by Griffiths. \cite{Griffiths12}

Equally intriguing is the case where electromagnetic momentum is nil in a charged system not only when it is static but also when the 
system is in motion. This happens in spite of the fact that electromagnetic energy is present in the system in either case, though it is 
even more baffling that the energy decreases when the system gains motion. We track down the presence of electromagnetic momentum in these 
cases that exists in the form of a continuous transport of electromagnetic energy from one part of the system to another, without any net 
gain of energy in the process. We also demonstrate how a charged system could comprise less energy after it has gathered motion from 
its initial state of rest. 
\section {An apparition of momentum without any apparent motion}
In order to comprehend the occurrence of momentum without motion we consider a specific example of two concentric spherical shells, one 
of radius $a$ having a surface charge density proportional to $\cos \theta$, resulting in an electric dipole 
moment ${\bf p}$, and the second sphere of radius $b$ having a surface current density proportional to 
$\sin \theta$, resulting in a magnetic dipole moment ${\bf m}$. The two spheres are oriented 
such that ${\bf p}$ and ${\bf m}$ are perpendicular to each other, which results in an electromagnetic linear 
momentum ${\bf  P}=\mu_{\rm o}{\bf m \times p}/(4\pi b^{3})$ in this otherwise static system. \cite{Romer95, Pugh67} The system could be called 
stationary because even a constant current implies movement of charges, however as far as the electric and magnetic fields are concerned, 
there is no temporal change and in that sense it is a static system.
%---------------------------------------------
\begin{figure}[ht]
\scalebox{0.75}{\includegraphics{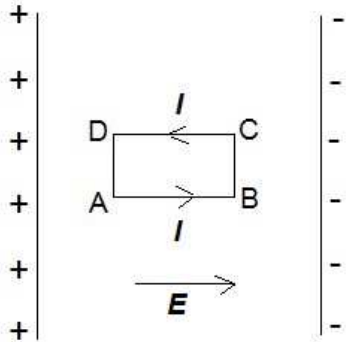}}
\caption{A current loop  ABCD, carrying current $I$, lying parallel to the constant electric field ${\bf E}$ of a charged 
parallel plate capacitor.}
\end{figure}
%---------------------------------------------
Now should one change the definition of electromagnetic momentum so as to make it zero, or 
is there some unknown ``hidden'' momentum that cancels the electromagnetic momentum to make it nil in a static system. In fact 
where does the electromagnetic momentum itself lie in such static systems? 
As the word ``momentum'' conjures up a vision of some kind of motion, where, if any, is 
such motion in the system corresponding to the electromagnetic momentum? 
%---------------------------------------------

To understand that, we first examine a simpler case of a 
small rectangular loop carrying a constant current $I=n\, e\, v$ and lying in a static electric field $\bf E$, assumed to be
constant and parallel to a side of the current loop. Here $n$ is the number of charges per unit length of the circuit,
$e$ is the electric charge of conducting charged particles, and $v$ is their drift velocity. 
On any small current element d$\bf l$ of the loop, the electric field does work
at a rate $n\, e\, v\, {\rm d}{\bf l \cdot E}= I\, {\rm d}{\bf l \cdot E}$. 
We notice that in one arm of the loop the electric field
$\bf E$ is doing a positive work while in the other parallel arm the work done is negative, and no work being done in the other two sides.
Thus while there may be a gain of energy in one arm of the loop, there is an equal loss in the opposite arm; effectively a continuous 
exchange of electrical energy takes place between opposite parts of the circuit because
of the presence of electrical field. Though in this process there is no net
change in the total energy content of the system, nevertheless there is a
linear momentum associated with this transport of energy across the loop. With respect to Fig. (1), 
the electric field pumps energy into arm AB and which is getting drained in arm CD, thus effectively electric field gives rise to a 
transport of electrical energy from side CD, where it disappears, to side AB where it appears. 
It is basically this transport of electric energy across the 
circuit that constitutes electromagnetic momentum in an otherwise static case.
%---------------------------------------------
\begin{figure}[ht]
\scalebox{0.8}{\includegraphics{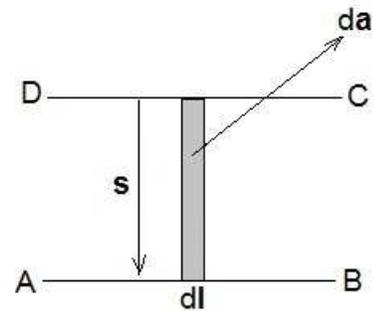}}
\caption{The geometry for calculating electromagnetic momentum due to transport of electric energy across the circuit.}
\end{figure}
%---------------------------------------------

The electromagnetic momentum can be calculated easily if we first consider a pair of current
elements placed symmetrically in the two arms of the loop (Fig. 2). The power being fed into the circuit element of length d$l$ in arm $AB$ 
through electric field is $I\: {\bf E} \cdot {\rm d}{\bf l}$, while a similar amount of power is being drained  
from an element d$l$ in arm $CD$. Effectively an energy 
$I {\bf E} \cdot {\rm d}{\bf l}$ is being transported per unit time along ${\bf s}$, the vector joining the  
current elements in side  $CD$ to that in $AB$, implying a momentum,
\begin{equation}
\label{eq:p31.1a}
{\bf E} \cdot {\rm d}{\bf l}\: {\bf s} \frac {I}{c^2}= {\bf E} \times {\rm d}{\bf a}\frac {I}{c^2}, 
\end{equation}
where d$\bf a$ is
the area across the loop contained between these two current elements. It is certainly not 
implied here that necessarily a one-to-one energy transfer takes place
between current elements placed symmetrically in two opposite sides of the circuit, 
this procedure, a simple book-keeping, is adopted merely to make the 
calculations easy by taking advantage of the symmetry of the system.
We integrate over the whole loop to find the
total linear momentum to be,  
\begin{equation}
\label{eq:p31.1b}
\frac {{\bf E} \times {\bf a}\,I}{c^2}= \frac{\bf E \times \bf m}{c^2}, 
\end{equation}
where ${\bf m=a}\,I$ is the magnetic
dipole moment of the current loop. It should be noted
that the electric current could be due to either positive or/and negative
charges, making no difference in the final result. We could replace the
rectangular shape of the loop with a polygon with larger number of sides and in the limit even
with a circular one without changing any of our essential arguments;
in fact we could have any irregular shaped loop as it can be always realized
by a superposition of a sufficiently large number of regular shaped loops.
Moreover $\bf E$ vector now need not necessarily be only in the plane of the
current loop. This therefore is a fairly general
result as long as the current loop is small enough for $\bf E$ to be
considered constant over its extent.

The system of course remains stationary
as there will be an equal and opposite momentum associated with the
stabilizing forces (irrespective of the detailed model of these forces of constraint) 
that keep the current in the loop everywhere constant (as we assumed it to begin with) in spite of the
effects of the electric field ${\bf E}$ on the loop (see Section V).

It is now straight forward to apply this result to the two spherical shells, \cite{Romer95, Pugh67} 
comprising a pair of electric and magnetic dipole, oriented perpendicular to each other. The magnetic dipole moment
$\bf m$ comprises a uniformly charged spherical shell of radius, say, $b$ rotating at a
constant rate$^{1}$. The external electric field of the electric dipole
moment $\bf p$ can be written as  
${\bf E}=[3({\bf p.\hat{r}}){\bf \hat{r}}-{\bf p}]/(4\pi\epsilon_{\rm o} r^{3})$.  \cite{Romer95, Pugh67} 
Ignoring the radial component of the field which does no work on 
conducting charges in the  spherical shell, we find the rest of
electric field component on the shell surface to be a constant
$-{\bf p}/(4\pi\epsilon_{\rm o} b^{3}$). Dividing the spherical shell into a
large number of circular current loops and then summing over them
we arrive at the total electromagnetic linear momentum of the system as
$\mu_{\rm o}{\bf m \times p}/(4\pi b^{3})$, the result obtained 
using the conventional formula for the electromagnetic field momentum.  \cite{Romer95, Pugh67} 
%---------------------------------------------
\section {The case of missing momentum in a moving charged system}
We shall now show that it is equally possible to have a charged system which has a nil electromagnetic momentum irrespective of whether the 
system is static or is in motion. Consider again the last example of a parallel plate capacitor, sans the current loop. 
However now the capacitor is moving parallel to the plate separation with a velocity ${\bf V}$ in the 
lab-frame ${\bf S}$ while ${\bf S}'$ is the rest frame of the charge (Fig. (2)).

Let $Q$ and $-Q$ be the total charges on the capacitor plates, each of area $A$, these quantities being invariants between ${\bf S}'$ and 
${\bf S}$. The electrostatic field, $E=Q/(A\epsilon_{\rm o})$, a constant in the region between the plates, is also the same in 
either frame. Similarly the mutual force of attraction between plates, $F=Q^{2}/(2 \epsilon_{\rm o} A)$, as well as the energy density, 
$\epsilon_{\rm o} E^2 /2 =Q^{2} /(2\epsilon_{\rm o} A^2)$, of the electrostatic field do not change between two frames. 

Let the separation between the two plates be $l$ in the lab-frame ${\bf S}$, then there will be an electric field energy 
$U = Q^{2} l/(2\epsilon_{\rm o}A)$, as calculated in ${\bf S}$. However, the magnetic field is zero throughout, 
therefore the field momentum is zero, which is rather perplexing since the system with an energy $U$ and moving with a velocity 
${\bf V}$ is expected to possess a momentum $U{\bf V}/c^2$ in the lab-frame ${\bf S}$. Why is the momentum nil for a moving system?
%---------------------------------------------
\begin{figure}[ht]
\scalebox{0.5}{\includegraphics{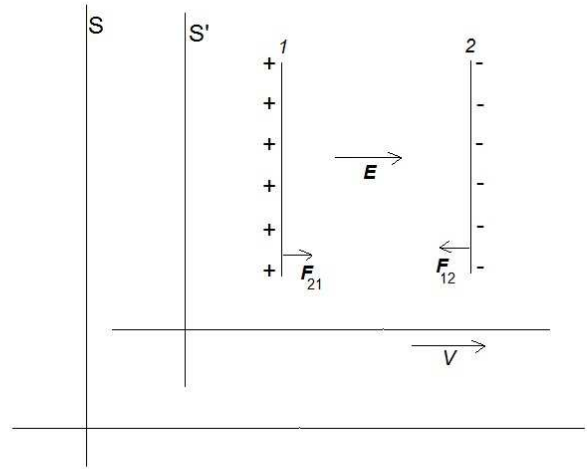}}
\caption{A charged parallel plate capacitor stationary in its rest-frame ${\bf S}'$ which is moving with a velocity ${\bf V}$ 
with respect to the lab-frame ${\bf S}$.}
\end{figure}
%---------------------------------------------

Actually due to the force of attraction between the plates, there is a negative 
momentum component (i.e., in a direction opposite to the
motion of the capacitor), due to the energy flow caused by the
force of attraction between the moving plates.
This excess momentum is over and above the usual relativistic momentum formula 
$U {\bf V}/c^{2}$. This
extra momentum is due to the fact that there is a continuous flow of energy {\em into} the system at its 
plate {\em 1} due to the 
work being done at a rate ${\bf F_{21}\cdot V}$, due to $\bf F_{21}$, the force
of attraction on plate {\em 1}.
At the same time an equal amount of energy is flowing {\em out} of the system due to $\bf F_{12}$ at 
plate {\em 2}. Although there is no net increase
of the system energy due to these forces, there is a continuous
transport of energy taking place from plate {\em 2} to plate {\em 1} through the force of attraction, 
which forms a part of the total electromagnetic momentum of the system. 
One can easily visualize the effect of work being done by these electromagnetic forces, 
if we consider the capacitor plates at this stage not clamped together 
and thus free to move under these forces of attraction. Then as seen in
the lab-frame ${\bf S}$, plate {\em 1} will gain a velocity higher than ${\bf V}$ while 
the velocity of plate {\em 2} will reduce below ${\bf V}$, implying a transport of energy 
at a rate $({Q^{2}}/{2\epsilon_{\rm o}A}){V}$, through electric attraction, across length $l$ of the capacitor system 
(i.e., from plate {\em 2} to plate {\em 1}). The resulting momentum is in a direction opposite to ${\bf V}$.

Therefore the total electromagnetic momentum of the system is,
\begin{equation}
\label{eq:p31.2a}
{\bf  P}_{\rm em}=U \frac{\bf V}{c^{2}}- \frac{Q^{2}}{2\epsilon_{\rm o}A}\frac{\bf V}{c^{2}}l\,=\,0.
\end{equation}
Thus we see that to get a consistent picture of the electromagnetic field momentum, 
one needs to track down and properly account for the contribution of the electromagnetic
forces during the motion of the system, the contributions that might not be so apparent in many cases. 
%----------------------------------------------------
\section{An intriguing case of energy decreasing with motion!}
The separations of the two plates, $l$ and $l'$ in frames ${\bf S}$ and ${\bf S}'$ are related, due to the Lorentz contraction,  
by $l=l'/\gamma$. Therefore the electric energies $U=Q^{2} l/(2\epsilon_{\rm o}A)$ and $U'=Q^{2} l'/(2\epsilon_{\rm o}A)$, obtained from 
the volume integrals of the energy density in ${\bf S}$ and ${\bf S}'$, imply $U=U'/\gamma$. This appears to be at variance with the usual 
relativistic transformation formulas where one expects the energy of the moving system  in lab-frame ${\bf S}$ to be instead higher by
a factor $\gamma$. How come the energy has decreased with motion?

To understand this, consider a case where the charged capacitor, initially at rest in the lab-frame ${\bf S}$ with a 
plate separation $l$ and energy $U=Q^{2} l/(2\epsilon_{\rm o}A)$, is given a push to acquire a velocity ${\bf V}$ so as to 
come to rest in ${\bf S}'$.  
In frame ${\bf S}'$ the capacitor was initially moving with a velocity ${-\bf V}$  with a plate 
separation $l/\gamma$ and energy $U=Q^{2} l/(2\epsilon_{\rm o}A\gamma)$. 
In lab-frame ${\bf S}$ both plates of the capacitor were given the push simultaneously,   
thus both plates continue to move with a velocity ${\bf V}$ with the same  
plate separation $l$ afterwards. In frame ${\bf S}'$, it will appear that the capacitor plates initially moving with  
velocity ${-\bf V}$ are being stopped so as to come to rest in ${\bf S}'$. However, like in the famous pole and barn problem, \cite{TW92}
the stopping of the plates will not be a simultaneous event in ${\bf S}'$, and the plate {\it 2} will come to rest before plate {\it 1} by a 
time interval $\Delta t'= \gamma l V/c^2$,
thus plate {\it 1} will continue to move with a velocity ${-\bf V}$ during the time $\Delta t'$ and will thus be a distance 
$\Delta l'=\gamma l V^2/c^2$ farther from plate {\em 2}, before it also comes to rest in ${\bf S}'$. Therefore the plate separation in 
${\bf S}'$ will become $(l/\gamma) + \Delta l' = l \gamma=l'$. Also during the time $\Delta t'$ the movement of plate {\it 1}  
will do work against the force of attraction on it and that will add to the energy of the system which will add to the electromagnetic 
field energy of the system $\Delta U= F_{21} \gamma l V^2/c^2$ making the net electromagnetic energy of the system in 
the rest frame ${\bf S}'$ as, 
\begin{equation}
\label{eq:p31.3a}
\frac{Q^{2} l}{2\epsilon_{\rm o}A\gamma}+ \frac{Q^{2} l\gamma}{2\epsilon_{\rm o}A}\frac{V^2}{c^2} = \frac{Q^{2} l\gamma}{2\epsilon_{\rm o}A}
= U \gamma. 
\end{equation}
Thus energy in the rest frame ${\bf S}'$ is $\gamma$ times higher than the energy in the lab-frame ${\bf S}$ with respect to which the 
capacitor system is moving with a velocity ${\bf V}$. This of course is consistent with the energy calculated from the standard formula for 
energy in electromagnetic fields in the system for each reference frame.
%------------------------------------------------
\section{Mechanical momentum due to the stabilizing forces}
There are also forces of stabilization in all such cases. We do not specify the ultimate nature of these forces 
as that may not have any relevance for the problem under discussion. All we know or care for is that these keep 
the system in equilibrium by providing stabilizing forces equal and opposite to any unbalanced electromagnetic forces in the system. 

For instance, in the case of the current loop of Section II, if a uniform current is not maintained using some stabilizing forces 
or constraints then a higher value of electric current in arm AB and a lower current in arm CD (due to the imposed electric 
field ${\bf E}$) will soon result in piling of positive charges on arm BC and a deficiency of similar charges in arm DA (or equivalently an   
accumulation of negative charges on arm DA), and this accruing will continue till the electric field giving rise to this non-uniform 
current is cancelled in the circuit. Of course in that case we will no longer find an electromagnetic momentum in the system, 
but that may not come as a surprise since we would not have an electric field either in the region of ${\bf m}$. 
Therefore in order to study the momentum arising from ${\bf E} \times {\bf m}$, we need to maintain a uniform electric current in the 
circuit using some constraints.  

One way to invoke such forces of constraint could be to adopt a simple model of current, 
with charges spread with equal spacing on a conveyor belt going around in a circuit like ABCD of Fig. (1), 
constrained to move with a uniform speed along its path so as to maintain a constant and uniform electric current. As the force on charges 
due to the imposed field ${\bf E}$ is parallel to the drift velocity ${\bf v}$ on one side of the circuit, the field does a positive 
work on that side and there will be an equal but negative work done on the opposite side of the circuit. 
Thus the agency, in order to keep a constant velocity of the belt and thereby maintain a uniform current, will gain 
energy on arm AB of the circuit but will impart energy on arm CD. Thus the agency drains energy on one side of the circuit and feeds equal amount 
of energy on the other side, thus a mechanical transport of energy from the side AB to the side CD is continuously taking 
place, and therefore a mechanical momentum exists in the circuit. Such has sometimes been termed as a ``hidden momentum'', this name 
is perhaps unfortunate \cite{Romer95, Griffiths12}, as it is  not any more hidden than the electromagnetic momentum itself which is 
also in the form of electromagnetic energy flow from one part of the system to another.  
As this momentum is from the transport of mechanical energy due to the forces of constraint that maintain a steady current in 
the system, in spite of the applied electric field, it would be termed better as ``mechanical momentum'' 
to distinguish it from the electromagnetic momentum. 
Another good example of the appearance of mechanical momentum in the system when electric current is constrained to be uniform across 
the circuit is given in Griffiths \cite{Griffiths12, 25}. The momentum due to the electromagnetic origin and that due to the mechanical cause 
are  equal and opposite and thus cancel each other making the net momentum (electromagnetic plus mechanical) of this static system nil. 
Thus we see that even though the total momentum of this static system is zero (that fits the name ``static''), the pure electromagnetic 
part of the momentum is not zero. We get into complications when we try to describe the {\em total} momentum of the system on a pure 
electromagnetic basis.

In the same way, the separation between the capacitor plates in Fig. (2) is kept {\em fixed} in a given frame, 
in spite of the mutual force of attraction 
between the plates, then there must be some forces of constraint, e.g.,  a non-conducting rod keeping the plates apart 
(or may be some clamps holding the plates fixed in their positions). Irrespective of the nature 
or details of these forces of constraint, without any ambiguity, on each plate they must be equal 
and opposite to the force of electrical attraction.
Therefore their contributions to the energy and momentum of the system must also be 
equal and opposite to that calculated above for the electrical forces.

For instance, there must be a mechanical momentum in the rod
holding the two capacitor plates apart. This is because as the system 
moves, as seen the lab-frame in ${\bf S}$, work is being done {\em on} the holding-rod by plate {\em 1} 
at a rate ${\bf F_{21}\cdot V}$, and an equal amount of work $-{\bf F_{12}\cdot V}$ is 
being done {\em by} the rod on plate {\em 2}. Although there is no net increase
of the energy of the rod due to this, there is a continuous
transport of energy taking place from the left-end of the rod to its
right-end, and this energy-flow through the rod is nothing but a 
mechanical momentum ${\bf P}_{me}=(Q^{2}{\bf V} l/(2\epsilon_{\rm o}A c^2)$. 

The total momentum of the system will be given by the momentum ${\bf P}_{em}$ in the electromagnetic fields 
(which could represent only the {\em pure} electromagnetic
part of a system) plus ${\bf P}_{me}$, the mechanical momentum associated with 
the forces of constraint as calculated above.
Therefore the {\em total} (electromagnetic plus  mechanical) momentum of the system is given by
\begin{equation}
\label{eq:p31.4a}
{\bf P}_{total}= {\bf  P}_{\rm em} + {\bf P}_{me} =U \frac{\bf V}{c^{2}}.
\end{equation}
The effect of such stabilizing forces has been shown to be essential for a successful explanation \cite{58} of the null results of 
the famous Trouton-Noble experiment. \cite{59}
%---------------------------------------------
\section{Discussion}
Through a couple of simple examples, we demonstrated above how the unbalanced electromagnetic forces in a system 
make additional contribution to the energy momentum of the system. The standard formulas for energy momentum of the electromagnetic 
fields of a system already contains all these electromagnetic contributions to the {\em electromagnetic} energy momentum. 
This brings us to the more general question raised by Romer: \cite{Romer95}
Are the conventional formulas for energy and momentum of
electromagnetic fields valid for all cases or do we need to change
over, at least in the case of ``bound fields,'' to some modified
definitions \cite{Rohrlich60}, which have  made their
appearance even in standard text books? \cite{1,2} To get an idea of these modified definitions, we consider a simple system in which an 
inertial frame, say ${\bf S}'$ can be found in which all charges are at rest. Then there is no magnetic field and only an electric field exists  
in frame ${\bf S}'$. Then energy $U'$ is given by the volume integral, 
\begin{equation}
\label{eq:p31.5}
U'=\frac{\epsilon_{\rm o}}{2}\int_{\rm V'}{E'^2}\:{\rm d}v'. 
\end{equation}
Then in the lab-frame ${\bf S}$, which is moving with a 
velocity $\bf V$ with respect to ${\bf S}'$, the electromagnetic field energy and momentum in the modified definitions \cite{2} are given by  
$U=\gamma U'$ and ${\bf P}=\gamma {\bf V} U'/c^2$. Accordingly field energy and momentum in ${\bf S}$ are given as, \cite{1}
\begin{equation}
\label{eq:p31.5a}
U=\frac{\epsilon_{\rm o}}{2}\gamma^2\int_{\rm V}(E^2-c^2B^2)\:{\rm d}v. 
\end{equation}
\begin{equation}
\label{eq:p31.5b}
{\bf  P}=\frac{\epsilon_{\rm o}}{2}\frac{\gamma^2\bf V}{c^2}\int_{\rm V}(E^2-c^2B^2)\:{\rm d}v. 
\end{equation}

Actually this question was
discussed previously at length \cite{15, Boyer85} where it is explicitly shown for the spherical model of a charged particle that 
when we consider the contributions of all  electromagnetic forces (which
may not always be so obvious, and that according to us is primarily the reason for so
much confusion in this subject) to the energy momentum of the considered
system we invariably arrive at the result derived ``independently'' from
conventional electromagnetic field energy-momentum formulas. Thereby a natural explanation for the famous but 
intriguing factor of 4/3 in the inertial electromagnetic mass of a charge, emerges without any ambiguity and there is
no compelling reason for adopting a change in the conventional formulas.
It is not even simply a matter of mere convenience or an individual's point of view
as sometimes advocated in the literature. \cite {Teukolsky96} One can put forward 
the following additional arguments in support of the conventional formulas vis-a-vis the 
modifications as referred to above.

1. The {\em total} energy-momentum of a charged system is the energy-momentum of
its electromagnetic parts plus the energy-momentum associated with 
the {\em other} (stabilizing, non-electromagnetic!) forces of constraint.
Since the electromagnetic  fields could represent the energy-momentum
associated with only the electromagnetic phenomenon 
(after all expressions for {\bf E} and {\bf B} make no reference and hence
contain no information about these ``other'' forces),
the energy-momentum associated with the stabilizing forces
could not be represented by these. Therefore a modification in definition of 
energy-momentum in electromagnetic fields to include contribution of these ``hidden'' 
non-electromagnetic forces is certainly not appropriate. 

2. In general, electromagnetic fields represent values at space-time events and these 
fields themselves are not assigned any velocities unlike a localized material 
particle. Therefore to choose one particular reference frame (a sort of ``rest-frame'' for the fields) on a preferential basis  
to calculate the volume integral of electromagnetic energy-momentum is not proper. In fact it will be 
against the very spirit of relativity.

3. Even if the modified definition might work for a single charged particle or when all the charged particles in the system being considered 
are moving together as a bunch with the same velocity vector,
so that they all could have a common rest-frame, such a modified definition does not work when 
there are more than one particle, with different particles having different velocity vectors with respect to each other 
as seen in any inertial frame. In such a case there can be no common rest-frame for the particles and their so-called bound fields. 
For each such particle the fields have to be treated separately. In other words, what  
parts of the electromagnetic fields at any event belong to what particle need to be first recognized and then  
broken into contributions of each such particle separately. This way the contribution to the 
electromagnetic field energy-momentum has to be calculated separately for each particle, as each particle contributes differently 
in different reference frames. This is so because each such contribution has to be evaluated with respect 
to the rest-frame of the individual particle as the Lorentz factor $\gamma$ enters into the definition of the such ``bound''
electromagnetic fields (viz. Eqs.~(\ref{eq:p31.5a}) and (\ref{eq:p31.5b})) and has to be then integrated over all particles. That means algebraic 
sum of the electric and magnetic fields (as components of 
vectors or a tensor) of all contributions at that event do not suffice to calculate energy-momentum contributions of electromagnetic fields.
In fact such a resultant field will not comply with the principle of 
linear superposition and from the field values in one inertial frame one cannot transform to another frame using the standard 
field transformation laws. One might wonder whether such a modified definition serves any useful purpose at all. 

4. To have two separate definitions for so-called ``free'' and ``bound'' fields is in itself an unnecessary complication where one single 
definition has been giving the correct and consistent results and been doing fine in all cases. 

5. If we do modify the definition of electromagnetic field momentum for such static cases (in the strict modified definition as say, 
in Ref. [\onlinecite {1}], so that the electromagnetic field momentum is zero (or at least different) in static cases, then we end up in 
bigger problems as we will have to modify definition of momentum even in mechanical cases so as to conserve momentum in electromechanical 
phenomenon. The momentum due to the electromagnetic origin and that due to the mechanical cause are  equal 
and opposite and thus cancel each other making the net momentum of this static system nil (as in Section II). 
Now if we do modify the electromagnetic 
momentum definition to make it zero (or even somewhat different from the standard definition) in a static system like we discussed here, 
then we will end up with a net momentum in a static system because a modification in the definition of the electromagnetic momentum will 
not change the contribution of the mechanical momentum, unless of course we modify the definition of the mechanical momentum formula too. 
That will open a Pandora's box. It is thus best to stick to the standard definition more so as no real contradiction results from their 
usage. 

In our opinion time has come to abandon these modified definitions as these do not seem to serve any useful purpose except creating 
confusion in the subject. It is only the standard definitions which are ``correct'' and should only be retained in text-books.
%---------------------------------------------
\section{Conclusions}
We examined a couple of simple electromagnetic systems, first one possessing momentum without any apparent motion and a second one 
with motion but without any momentum. We showed that this mysterious electromagnetic momenta exists in the form of a transport of 
electrical energy taking place from one part of the system to another, all {\em within} the system. We resolved these puzzling, 
apparent paradoxes, within the 
conventional formulation of the electromagnetic field energy-momentum, thereby obviating the need for modifications in the definition 
of standard formulation that have been advocated in the literature. We demonstrated that the standard formulation suffices to explain all 
electromagnetic phenomena and that it is the standard formulation only which yields result consistent with the rest of physics, 
especially with the special theory of relativity,.
%--------------------
{}
\end{document}